\documentclass[12pt]{article}
\renewcommand\({\left(}
\renewcommand\){\right)}
\renewcommand\[{\left[}
\renewcommand\]{\right]}

\newcommand\eq[1]{Eq.~(\ref{#1})}
\newcommand\eqs[2]{Eqs.~(\ref{#1}) and (\ref{#2})}

\newcommand\ee{\end{equation}}
\newcommand\be{\begin{equation}}
\newcommand\eea{\end{eqnarray}}
\newcommand\bea{\begin{eqnarray}}

\newcommand\GeV{\,\mbox{GeV}}

\newcommand\mpl{M_{\rm P}}

\newcommand\lsim{\mathrel{\rlap{\lower4pt\hbox{\hskip1pt$\sim$}}
    \raise1pt\hbox{$<$}}}
\newcommand\gsim{\mathrel{\rlap{\lower4pt\hbox{\hskip1pt$\sim$}}
    \raise1pt\hbox{$>$}}}

\newcommand\diff{\mbox d}

\newcommand\calp{{\cal P}}
\newcommand\calr{{\cal R}}
\newcommand\calpr{{\calp_\calr}}

\newcommand\bfk{{\bf k}}

\newcommand\bfx{{\bf x}}

\newcommand\sub[1]{_{\rm #1}}

\begin{document}
\date{}
\title{The failure of cosmological perturbation theory in the new ekpyrotic 
and cyclic ekpyrotic scenarios}

\author{David H.~Lyth
\\
\\
Physics Department, Lancaster University, Lancaster LA1 4YB,  U.K.}
\sloppy
\maketitle

\begin{abstract}\noindent
Cosmological perturbation theory fails in the
new ekpyrotic and cyclic ekpyrotic  scenarios,
 before the scale factor of the Universe reaches
zero. As a result, a recently-proposed  recipe for evolving the curvature 
perturbation through the bounce in these scenarios cannot be justified.
\end{abstract}

 \paragraph{Introduction}
It is now clear that the dominant cause of large-scale
structure in the Universe
is a perturbation in the spatial curvature \cite{book}.
 The perturbation exists
already when cosmological scales are approaching horizon entry, at which stage
it has an almost flat spectrum. It  presumably  originates
during an early era of superluminal evolution, defined as one in which comoving
scales are leaving the horizon, at the beginning of which the entire
observable Universe is inside the horizon.
During superluminal evolution the curvature of the Universe in Hubble
units is decreasing, explaining naturally the  near-flatness of
the observable Universe.

The usual hypothesis concerning the 
 superluminal  era is that it is one of expansion, or in other words
an era of  inflation, with slowly varying Hubble parameter $H$.
During inflation, each canonically-normalized light scalar field
acquires a perturbation with a flat spectrum, and this provides
a natural origin for the observed curvature perturbation.
The curvature perturbation might be generated directly during inflation
from the perturbation in the slowly-rolling inflaton field
\cite{book},
or indirectly  \cite{p01david} by a different field
(curvaton) whose isocurvature
density perturbation is converted into a curvature perturbation
during some radiation-dominated era after inflation.

The  alternative to inflation is an era of  superluminal collapse
(negative  Hubble parameter)
giving way at some stage to expansion (positive Hubble
parameter). Unless  spatial curvature becomes significant near the bounce,
the  theory describing  the bounce
cannot be standard field theory with Einstein gravity, corresponding
to an action of the form
\be
S= \int \diff^4x \sqrt{-g} \[ \frac12\mpl^2 R 
- \frac12 g^{\mu\nu} G_{IJ}(\phi^K) \partial_\mu\phi^I
\partial_\nu \phi^J - V(\phi^K) \]
\label{standard}
\,.
\ee
The reason is that the energy density $\rho$ and pressure $P$ in
such a theory satisfies the null energy condition
$\rho+P >0$, which implies that the energy density
 decreases monotonically. Through the Friedmann equation, this implies
 that the Hubble parameter decreases
monotonically in the absence of spatial curvature, forbidding any bounce.
The bounce might be described by a 
non-standard four-dimensional field theory, but more plausibly would
involve extra dimensions and/or string theory.

\paragraph*{The pre-big-bang scenario}
Until recently, only one type of superluminal collapse has been considered,
and the scenario involving it is termed simply the
 pre-big-bang scenario \cite{pbb,clw}. 
In this scenario the action is supposed to have
the standard form \eq{standard}
 during
both collapse and expansion,
up to some maximum energy density $\rho\sub b $. This is  the bounce energy
density, which 
is supposed to correspond to the string scale and is at or below
the Planck scale $\rho\sub P=\mpl^4 = (2\times 10^{18}\GeV)^4$.
During pre-big-bang collapse,  the energy density
of the Universe comes from a scalar field $\phi$ with no potential.
Taking time $t$ as negative this leads to scale factor
$a\propto |t|^\frac13$ corresponding to $H\equiv \dot a/a =1/3t$.
The same result is obtained if there are several such fields, each giving
a fixed fraction of the energy density.

\paragraph*{The ekpyrotic scenario}
Recently a different mechanism for superluminal collapse has been 
proposed, termed the ekpyrotic mechanism \cite{ek,ek3}. It invokes an extra
dimension in which two branes are attracted  towards each other.
After integrating out the extra dimension one obtains superluminal
four-dimensional collapse, in which the 
 the energy density and pressure are   dominated
by a canonically-normalized field $\phi$ which corresponds to the 
distance between the branes. The action is of the standard form
\eq{standard}, 
\be
S= \int \diff^4x \sqrt{-g} \[ \frac12\mpl^2 R 
- \frac12 g^{\mu\nu} \partial_\mu\phi
\partial_\nu \phi - V(\phi) + \cdots \]
\label{standardek}
\,,
\ee
the remaining terms having negligible effect on the evolution
of $\phi$.

From some very early time, until after cosmological scales have left the 
horizon, the  potential during collapse is taken to be
\be
V\simeq -V_0\exp\(-\sqrt\frac2 p \frac\phi\mpl\) 
\,
\label{5}
\ee
with $p\ll 1$. It is supposed that  $\dot\phi$ starts  out from a sufficiently
small value, leading 
 to power-law collapse $a\propto |t|^p$ in which the Hubble parameter
and energy density are related by the 
 Friedmann equation
\bea
3\mpl^2H^2 &=& \rho \label{1a}\\
& =& \frac12 \dot\phi^2 + V(\phi)
\label{1b}
\,.
\eea
The energy density increases during collapse, but it is always small
compared with $|V|$ by 
 virtue of the condition $p\ll 1$.

Three versions of the ekpyrotic scenario
have been proposed, in references  \cite{ek}, \cite{ek3} and
\cite{cyclic}, which we refer to as respectively the old, new and cyclic
ekpyrotic scenarios.  In
 each of them it is claimed 
that the observed curvature perturbation can be generated by the quantum
fluctuation of the field $\phi$. In an earlier note \cite{first}
we demonstrated
that this claim is invalid in the case of the old ekpyrotic scenario,
as described in \cite{ek}.
The  purpose
of  the present note is to show that the claim is also invalid in the
case of the new ekpyrotic scenario as described in reference \cite{ek3}, 
because in that reference
linear  cosmological perturbation theory is being invoked 
in a  regime where it is not valid. Since the cyclic ekpyrotic Universe
 \cite{cyclic} invokes the same mechanism for producing the density
perturbation, the same statement applies to that scenario also.

\paragraph*{The comoving curvature perturbation}

To define the cosmological perturbations one has to choose a gauge
(coordinate system) which defines a threading and slicing of spacetime.
Let us denote a generic perturbation by the symbol $g$. 
At each moment of time, it  
is Fourier-expanded in comoving coordinates
\be
g(\bfx) = (2\pi)^{-\frac32} \int g(\bfk)
e^{i\bfk\cdot\bfx} \diff^3 \bfk
\label{16x}
\,,
\ee
with  $k/a$ the wavenumber. 
 Its spectrum is defined by \cite{areview,treview,book}
\be
\langle g(\bfk) g^*(\bfk') \rangle = 2\pi^2 k^{-3} \delta^3
(\bfk-\bfk') \calp_g(k)
 \label{16}
\,,
\ee
where the brackets denote an  ensemble average.
(On the usual supposition that perturbations originate as quantum fluctuations,
the brackets denote the quantum expectation value.)
The formal expectation value 
of $g^2(\bfx)$ at  any 
point in space is 
\be
\langle g^2(\bfx) \rangle =\int^\infty_0 \calp_g(k) \diff (\ln k)
\,,
\ee
and $\calp_g^\frac12$ is 
the  the typical magnitude of a fluctuation in $g(\bfx)$
with size of order $1/k$.
The perturbation has a time-dependence, which is inherited by the spectrum,
\be
\calp_g^\frac12(\bfk,t) = {\,\rm const\,} |g(\bfk,t)|
\label{specev}
\,.
\ee

To define the spatial curvature perturbation we need a slicing of spacetime,
which is best  taken as the one orthogonal to comoving worldlines,
the comoving slicing. Its line element may be written
\be
\diff\ell^2 = a^2(t) (1+2\calr) \delta_{ij} \diff x^i \diff x^j
\,,
\label{calrdef}
\ee
and $\calr$ defines the curvature perturbation.\footnote
{This is the $\phi\sub m$ of \cite{bardeen}.
On the super-horizon scales of interest the comoving slicing is
practically coincident with the slicing of uniform energy density
which can equally well be used. The corresponding perturbation
 is the  $\zeta$ of \cite{bst}.}
This is the best choice because under very general circumstances
$\calr$ is time-independent 
on super-horizon scales. As a result, the curvature perturbation that
is generated in the very early Universe should be the one that is observed
as cosmological scales start to enter the horizon. 

There is also a slicing with no curvature perturbation, the
flat slicing.
 If the displacement from the flat slicing to the comoving one
is $\delta t(\bfx,t)$ the curvature perturbation on the latter is
$\calr = |H|\delta t $.
(A similar  formula actually gives the curvature perturbation on 
any slicing.) 
Whenever the energy density and pressure are dominated by single
field $\phi$, this field is uniform on comoving slices \cite{bst}. It follows 
\cite{ks,sasaki} that $\calr =- |H|\delta\phi/\dot\phi$,
where $\delta\phi$ is defined
 on the spatially flat slicing. (It is the `gauge-invariant' field
perturbation of \cite{sasaki}.) 


\paragraph*{The curvature perturbation in the pre-big-bang scenario}
The spectrum of the curvature perturbation during superluminal evolution
is found by a standard method \cite{book}. In the 
 pre-big-bang case one finds  \cite{pbb,clw},
well after the epoch $aH=k$ of
horizon exit,
\be
\calpr^\frac12 =  \frac1{2\sqrt 6 \pi^\frac32} 
 \(\frac{|H |}{\mpl} \) 
\(\frac k{a |H |} \)^\frac 3 2
\ln\( \frac {a|H|} k \)
\,.
\label{rspecpbb}
\ee
This spectrum is  time-independent except for the logarithm,
and corresponds to spectral index $n=4$. 


Assuming that the curvature perturbation is continuous across the bounce
it has to be generated in the pre-big-bang scenario after the bounce
by some curvaton field.
To possess  a flat spectrum in the face of the 
rapidly-varying $H$, this field must have \cite{davidref,clw,kari,p01david}
 a suitable initial condition,  and 
a  suitable coupling to $\phi$.
An alternative proposal might be to 
generate the curvature perturbation (or conceivably a 
curvaton field perturbation) during the bounce, but such a thing has
never been suggested let alone a mechanism for achieving it.

\paragraph*{The curvature perturbation in the old ekpyrotic scenario}
In the old  ekpyrotic scenario \cite{ek},
it is supposed that the brane collision
corresponds to some finite field value $\phi $, corresponding
to energy density $\rho $. After the collision the Universe
is expanding with the same initial energy density and there is no $\phi$
field.
In this scenario the spectrum of the
curvature perturbation during collapse can
be calculated in exactly the same way as for the pre-big-bang, but taking
into account the exponential potential.\footnote
{In \cite{ek} it is stated that just before collision the potential
climbs abruptly back to zero, but this has a negligible effect on the
result.}
Well after horizon exit, to leading order in $k/aH$, the spectrum
\cite{first}
is time-independent with spectral index $n\simeq 3$. 
This result is specific to the exponential potential \eq{5}
but there is no reason to think that a different potential would
give the necessary flat spectrum and in any case something like
\eq{5} is motivated by string theory. The conclusion therefore is that
 in the old ekpyrotic scenario, the observed curvature perturbation has to 
be generated either at the bounce or subsequently by a curvaton
field.

This conclusion is different from the one reached in
the  original ekpyrotic paper \cite{ek} (see also \cite{py}).
That  paper ignored
the effect of the metric perturbation on the field evolution
(gravitational back-reaction), and arrived at the
 incorrect conclusion that the comoving curvature
perturbation generated during collapse has a flat spectrum,
 so that it could
be the observed curvature perturbation.\footnote
{The words `curvature perturbation' do not appear in \cite{ek},
which uses the formalism of Guth and Pi \cite{gp} and Olson \cite{olson}.
However, as pointed out in \cite{first},
that formalism is  equivalent to the usual one, because of the formula
 $\calr=|H|\delta t$ where $\delta t$ is the time delay introduced by 
Guth and Pi. We emphasize that the equivalence is
complete  provided that $\calr$
 is correctly evaluated. The approximation 
 that $\delta t$ is constant,
made by Guth and Pi, is inessential though it is valid during
inflation in the slow-roll approximation. Usually,  $\calr$ is constant
on super-horizon scales, which requires $\delta t\propto 1/H$.}

\paragraph*{The  new 
and cyclic  ekpyrotic scenarios}
In the  new  \cite{ek3}
ekpyrotic scenario
it  is supposed, on the basis of string theory,
that the brane collision corresponds
 to $\phi=-\infty$. There  are 
 two stages of collapse. In the first stage,
which ends only after cosmological scales have left the horizon,
the potential is given by \eq{5} as in the old ekpyrotic scenario.
In the second stage,  the potential climbs towards zero  and 
becomes negligible compared with the kinetic term
so that the evolution of perturbations is the same as in the 
pre-big-bang scenario. This stage ends when
 $\phi=-\infty$,
 corresponding  to 
 infinite energy density,
infinite Hubble parameter and zero scale factor $a$. 
The branes then collide and the expansion of the 
Universe begins,  but in contrast with
the old ekpyrotic scenario the branes move apart 
after the collision so that the
 field $\phi$ continues to exist, though with a possibly different
potential.
The potential is at least initially negligible so that the
 contribution of $\phi$  to the energy density falls like
$a^{-6}$. In addition there is  the
energy density of radiation produced by
the brane collision, which 
falls like $a^{-4}$ and eventually dominates leading to the 
usual Hot Big Bang. 

The cyclic ekpyrotic scenario
\cite{cyclic} is the same, except that the potential of  $\phi$
is  the same during expansion as during collapse, and this potential
asymptotes at late times to a nonzero positive constant. The latter
 feature
gives rise to a slowly-increasing cosmological constant and eventually
to the re-collapse of the Universe and therefore to a continuous cycle of
expansion and collapse. None of this matters for the present purpose,
because the 
mechanism for generating the observed curvature perturbation proposed
in \cite{cyclic} is the same
as the one proposed in \cite{ek3}.
Accordingly we refer from now on to the `new' ekpyrotic scenario, taking
it as read that the statements apply also to the cyclic ekpyrotic scenario.

The general idea of the new ekpyrotic scenario,
 developed in an earlier related paper
\cite{kosst}, is that the underlying five-dimensional theory
is non-singular as one goes through the bounce, so that it may in some
sense be
 permissible to allow the 
 four-dimensional scale factor to go through zero. The  standard field
theory action \eq{standard} is supposed in some sense to remain valid
all the way down to $a=0$.
It is of course well-known that 
 \eq{standard}
cannot be literally  valid  on time- and distance scales below the
Planck scale, because the quantum fluctuation on such scales 
is out of control. For instance,  on the distance 
scale $\mpl^{-1}$ the vacuum fluctuation in the energy
density corresponding to the field gradients
is of order $\mpl^4$, which  is big enough in relation to the 
distance scale to form black holes (the spacetime foam).
In particular, the theory cannot be taken literally when the Hubble
time  falls below the Planck scale, corresponding to 
$H\gsim \mpl$ and $\rho\gsim \mpl^4$. 
This problem is clearly recognized in \cite{kosst} (``there is concern that
the quantum fluctuations become uncontrolled at $t=0$'').

Nevertheless, the action of \eq{standardek} is used in \cite{ek3} to 
calculate a  curvature perturbation which, after the 
bounce, has the flat spectrum required by observation. 
The purpose of this note is to point out that the calculation is invalid,
because it invokes linear cosmological perturbation theory outside its
regime of validity.

\paragraph{The calculation of reference \cite{ek3}}
The calculation in \cite{ek3} starts with the
Bardeen potential \cite{bardeen}, which is usually  defined as the spatial
curvature perturbation in the conformal Newtonian gauge;
\be
\diff s^2 = a^2(t) \( - (1+2\Psi) \diff\tau^2+
 (1-2\Phi)\delta_{ij}\diff x^i \diff x^j \)
\label{newtonian}
\,.
\ee
Alternatively, it may be defined in terms of the
 density contrast $\delta\equiv\delta\rho/\rho$ defined on 
 comoving slices (the Bardeen \cite{bardeen} variable $\epsilon\sub m$),
\be
\delta = -\frac 23 \(\frac k{aH} \)^2 \Phi
\label{deltaofPhi}
\,.
\ee
The definitions are equivalent within linear perturbation theory,
but the first one requires $|\Phi|\ll 1$ which is quite restrictive
\cite{bardeen}
and the second one is preferable.

The spectrum of $\Phi$ during the first stage of
collapse may in principle  be calculated
in exactly the same way as the spectrum of $\calr$, using the
evolution equation for the field perturbation $\delta\phi\sub N$
and imposing the same condition of an initial vacuum state.
(A simpler technique \cite{robert,ek3} is  to 
consider the evolution of $\Phi$, using its relation with
 $\delta\phi\sub N$ only to impose the initial vacuum condition.
A third method would be to deduce $\calp_\Phi$ from the
(slowly-varying) $\calpr$, using 
\eq{specev} with \eq{req} below.)
It turns out that during this stage the   evolution of 
 $\delta\phi\sub N$ 
is not  much affected by gravitational back-reaction 
 and one finds
\be
\calp_\Phi^\frac12 = \frac1{2\pi} \sqrt\frac1{2p}  
\frac{H  }{\mpl} \(
\frac{a  H  }k \) ^p
\label{phispec}
\,.
\ee
The last factor gives the effect of
 gravitational back-reaction. 

The idea behind the proposal of \cite{ek3}
is that the flat spectrum \eq{phispec} of the Bardeen potential 
may be converted
at the bounce into a flat spectrum for $\calr$, through the 
 coupled evolution equations of 
Einstein gravity. In these equations, anisotropic stress 
may presumably be ignored on the super-horizon scales of interest.
 The coupled equations are then \cite{areview,treview,book}
\bea
-\calr &=& \frac23 \frac{|H|^{-1}\dot\Phi +\Phi}{1+w} +\Phi 
\label{phieq}\\
|H|^{-1} \dot\calr &=& -\frac{\delta P}{\rho + P}
\label{req1} 
\,,
\eea
where $\delta P$ is defined on the comoving slicing.
The non-adiabatic pressure perturbation
 $\delta P\sub{nad}\equiv \delta P- (\dot P/\dot \rho) \delta 
\rho$ is also ignored on on super-horizon scales, which is appropriate
in a scenario where one is trying to generate the curvature perturbation 
without a curvaton \cite{p01david,kari} field. Then \eq{req1} becomes
\be
|H|^{-1} \dot\calr = -\frac{c_s^2}{1+w} \delta
\equiv  \frac32 \frac{c_s^2}{1+w} \(\frac k{aH} \)^2 \Phi
\label{req} 
\,,
\ee
where $c_s^2\equiv \dot P/\dot \rho$.

\eqs{phieq}{req} form a closed system with two independent solutions.
{}From \eq{req} it follows  that {\em $\calr$ is practically
constant provided that $c_s^2|\delta|\ll (1+w)|\calr|$}. 
This condition is very well
satisfied
in the expanding Universe for any 
reasonable initial condition and any reasonable 
behavior of $a$ and $w$ \cite{treview,book}. It is satisfied during
the first stage of ekpyrotic collapse and marginally violated during
pre-big-bang collapse. 


During both stages of ekpyrotic collapse the left hand side of
\eq{phieq} is negligible leading to $\Phi\propto H/a$. 
Using this result and \eqs{specev}{phispec},
the spectrum of $\delta$ in the limit $t\to 0$
when the potential has become negligible is given by
$\calp_\delta^\frac12 \to  A k^{1+p} $
with $A$ a constant. Using \eqs{specev}{phieq} this implies that 
the spectrum of $\calr$ is logarithmically divergent,
$6\calpr^\frac12 \to  \calp_\delta^\frac12 \ln |t|$.

Now we come to the  crucial point. Because $\calr$ is divergent,
the authors of  \cite{ek3} focus on  $\delta$.
Combining \eqs{req}{phieq} gives an equation of the form
 \cite{bardeen,lyth85}
\be
\ddot \delta + A(t) \dot\delta + B(t) \delta = 0
\,.
\label{secondorder}
\ee
 Assuming that this equation remains valid when $\calr$
goes infinite, the authors of \cite{ek3} impose  a scale-independent
continuity condition on $\delta(t)$ across $t=0$. This leads at
$t>0$ to a  Bardeen potential with the 
mild scale dependence given
by \eq{phispec}, 
and to  a curvature perturbation which by the time of 
 radiation domination has the same scale-dependence, which can agree with
observation. 

\paragraph{The failure of linear cosmological perturbation theory}
We will now show that 
 the analysis just described
is invalid,  by establishing the
following contradiction:  an {\em assumption} of the 
analysis is  that the  Universe has 
practically  uniform energy density and practically uniform (actually zero)
 spatial curvature,
so that the equations of linear cosmological perturbation 
theory and in particular \eq{secondorder} can be derived,
 yet a {\em conclusion}
of the analysis is that the above assumption is violated as the 
singularity is approached. 
To be more precise,  we shall demonstrate  that for every choice of slicing
used to define the purportedly linear theory, either  the density
contrast or   the curvature perturbation   blows up
as the singularity is approached (or both do), 
thus invalidating the linear theory
with that choice. 

The demonstration is very simple. We have seen already that on comoving
slices, the curvature perturbation $\calr$ blows up.
Now consider a
 generic slicing,
separated from the comoving one by time displacement $\delta t(\bfx,t)$.
On such a slicing, the density contrast
 $\delta\sub s$ and the curvature
perturbation $\calr\sub s$ (defined as in \eq{calrdef}) are
given by  \cite{bardeen,book}
\bea
\delta\sub s &=&  \delta - 3(1+w) |H| \delta t \nonumber\\
\calr\sub s &=& \calr + |H| \delta t
\label{gaugechange}
\,.
\eea
As the singularity is approached, $1+w$ approaches 
the nonzero value $2$, $\delta$ approaches a constant and $\calr$
diverges. Therefore, no choice of $\delta t$ can keep both $\delta\sub s$
and $\calr\sub s$ finite as the singularity is approached. 

One might think that the divergence of the curvature perturbation 
on comoving slices is by itself enough to invalidate the use
of \eq{secondorder} since the relevant  quantity $\delta$
has been defined as the density contrast on comoving slices. 
This definition obviously fails when the curvature of the slices diverges,
and moreover the derivation of \eq{secondorder} uses  \eqs{phieq}{req}
which apparently require $\calr$ to be small. However, within linear theory
quantities defined in a given gauge can always be redefined in terms of
quantities in any other gauge. (Recall that we gave an explicit example of
this in the case of the Bardeen potential $\Phi$.)
Therefore, if linear theory were valid in just one gauge, it would be 
possible to redefine $\delta$ in terms of quantities in that
gauge, and then \eq{secondorder} would be valid.

\paragraph*{Conclusion} 

It is   difficult to generate the observed
curvature perturbation in the pre-big-bang and ekpyrotic scenarios.
It  cannot be generated
during collapse and there is no known mechanism for generating it
at the bounce either. Subsequent generation by a suitable curvaton
field is possible, but 
the curvaton must have a suitable non-canonical
kinetic term and a suitable initial value.
All this is likely to hold in any scenario where inflation
is replaced by superluminal collapse, because it stems from
 the fact that the Hubble parameter is rapidly varying.
By contrast the curvature perturbation is easily generated 
in the inflation  scenario, either directly by the inflaton field
or indirectly by a curvaton field.

\paragraph*{Acknowledgement}
I thank David Wands for many helpful remarks about the evolution of 
the curvature perturbation. I have benefited also from correspondence with
Jai-chan Hwang, Robert Brandenberger, Ruth Durrer  and Gabrielle Veneziano.


\newcommand\pl[3]{Phys.\ Lett.\ {\bf #1}  (#3) #2}
\newcommand\np[3]{Nucl.\ Phys.\ {\bf #1}  (#3) #2}
\newcommand\pr[3]{Phys.\ Rep.\ {\bf #1}  (#3) #2}
\newcommand\prl[3]{Phys.\ Rev.\ Lett.\ {\bf #1}  (#3)  #2}
\newcommand\prd[3]{Phys.\ Rev.\ D{\bf #1}  (#3) #2}
\newcommand\ptp[3]{Prog.\ Theor.\ Phys.\ {\bf #1}  (#3)  #2 }
\newcommand\rpp[3]{Rep.\ on Prog.\ in Phys.\ {\bf #1} (#3) #2}
\newcommand\jhep[2]{JHEP #1 (#2)}
\newcommand\grg[3]{Gen.\ Rel.\ Grav.\ {\bf #1}  (#3) #2}
\newcommand\mnras[3]{MNRAS {\bf #1}   (#3) #2}
\newcommand\apj[3]{Astrophys.\ J.\ Lett.\ {\bf #1}  (#3) #2}
\newcommand\apjl[3]{Astrophys.\ J.\ Lett.\ {\bf #1}  (#3) #2}

\end{document}